# Linearized frequency domain Landau-Lifshitz-Gilbert equation formulation

Zhuonan Lin[1,2,3] and Vitaliy Lomakin[1,2]

[1] Department of Electrical and Computer Engineering, University of California San Diego, La Jolla, California 92093, USA
[2] Center for Memory and Recording Research, La Jolla, California 92093, USA
[3] Materials Science and Engineering Program, University of California San Diego, La Jolla, California 92093, USA

(*Electronic mail: vlomakin@eng.ucsd.edu)

(Dated: 3 October 2022)

We present a general finite element linearized Landau-Lifshitz-Gilbert equation (LLGE) solver for magnetic systems under weak time-harmonic excitation field. The linearized LLGE is obtained by assuming a small deviation around the equilibrium state of the magnetic system. Inserting such expansion into LLGE and keeping only first order terms gives the linearized LLGE, which gives a frequency domain solution for the complex magnetization amplitudes under an external time-harmonic applied field of a given frequency. We solve the linear system with an iterative solver using generalized minimal residual method. We construct a preconditioner matrix to effectively solve the linear system. The validity, effectiveness, speed, and scalability of the linear solver are demonstrated via numerical examples.

## I. INTRODUCTION

The magnetization dynamics is described by the Landau-Lifshitz-Gilbert equation (LLGE). LLGE is a time dependent non-linear equation, and it describes the magnetization dynamics in both linear and non-linear regimes. Solving LLGE can be computationally expensive as it requires obtaining solutions at many time steps and it is often numerically stiff, which may require either small time steps or assisting linear solvers. Under weak dynamics excitations, however, the magnetization dynamic response may be linear. Examples of such systems are spin wave excitations under weak time-dependent applied fields or the initial dynamics under spin torque excitations[1,2]. In such cases, LLGE can be linearized in terms of a weak magnetization deviation around the equilibrium state. Such linearization has been used to obtain solutions in terms of the eigenstate representations[3] and it can be used to obtain solutions even for non-linear problems[4–6].

Here, we present a linear frequency domain LLGE (FD-LLGE) solver. The FD-LLGE solver provides a linearized magnetization solution as a response to a dynamic excitation, e.g., the applied field, which is a characterized by a given frequency. Considering a given frequency allows formulating a time independent linear equation for a complex magnetization amplitude, which can, then, be used to provide linear time domain solutions. The FD-LLGE solver is developed based on the finite element based micromagnetic simulator FastMag[7]. The linear system is solved effectively with an iterative solver. The number of iterations is significantly reduced using a linear preconditioner. Solving the FD-LLGE is much more efficient and provides a more physically insight than solving the original LLGE.

## II. FORMULATION

The magnetization dynamics is described by the LLGE, which is, in its implicit form, is written as

$$\frac{\partial \mathbf{m}}{\partial t} = -\gamma \mathbf{m} \times (\mathbf{H}_{eff} + \mathbf{H}_a) + \alpha \mathbf{m} \times \frac{\partial \mathbf{m}}{\partial t}, \quad (1)$$

where $\mathbf{m}$ is the normalized magnetization, $\gamma$ is the gyromagnetic ratio, $\alpha$ is the damping constant. The term $\mathbf{H}_a = \mathbf{H}_{0a} + \mathbf{h}_a$ is the applied field containing a static component $\mathbf{H}_{0a}$ and a dynamic component $\mathbf{h}_a$, which is a time harmonic excitation at a circular frequency $\omega$, i.e., it can be written as

$$\mathbf{h}_a = \text{Re}\{\tilde{\mathbf{h}}_a(\mathbf{r})e^{j\omega t}\}, \quad (2)$$

where $\tilde{\mathbf{h}}_a$ is the complex amplitude of the exciting magnetic field, which represents its magnitude and phase. The term $\mathbf{H}_{eff}$ in LLGE (1) is the effective field. The effective field is a function of $\mathbf{m}$, and it is composed of several components, including the magnetostatic field $\mathbf{H}_{ms}$, exchange field $\mathbf{H}_{ex}$, and anisotropy field (assumed uniaxial) $\mathbf{H}_{an}$:

$$\begin{aligned}
\mathbf{H}_{eff} &= \mathbf{H}_{ms} + \mathbf{H}_{ex} + \mathbf{H}_{an} \equiv C\mathbf{m}, \\
\mathbf{H}_{ms} &= M_s \nabla \int \frac{\nabla \cdot \mathbf{m}}{|\mathbf{r} - \mathbf{r}'|} d\mathbf{r}', \\
\mathbf{H}_{ex} &= \frac{2A}{M_s} \nabla^2 \mathbf{m}, \\
\mathbf{H}_{an} &= H_K (\hat{\mathbf{k}} \cdot \mathbf{m})\hat{\mathbf{k}}.
\end{aligned} \quad (3)$$

Here, $M_S$ is the saturation magnetization, $A$ is the exchange constant, $H_K$ is the anisotropy field, and $\hat{\mathbf{k}}$ is the uniaxial anisotropy axis direction. The effective field is linear in $\mathbf{m}$ and, therefore, $C$ is the linear field operator that is independent of $\mathbf{m}$.

The LLGE (1) is non-linear in $\mathbf{m}$ due to the presence of the cross products and it describes the magnetization dynamics in a broad range of situations, including linear and non-linear effects. In various cases, however, the general LLGE can be linearized. Such a linearization is allowed when the magnetization varies insignificantly around its equilibrium state. Weak magnetization variations can be due to weak excitations, e.g., by weak applied fields or by spin transfer torque (STT). Even in the general non-linear cases, the initial dynamics that contains important information about the system behavior can be characterized by the linearization[3,4].

We present a framework that uses a linearized LLGE to study the magnetization dynamics. We first present a linearized time domain LLGE. Then, we construct a FD-LLGE to study small oscillations around the equilibrium state driven by time-harmonic excitations.

We seek a solution for small magnetization deviation $\mathbf{v}$ around the equilibrium state such that

$$\mathbf{m} = \mathbf{m}_0 + \mathbf{v} . \qquad (4)$$

Here, $\mathbf{m}_0$ is the equilibrium magnetization state for the system without a dynamic excitation, which is given by the Brown condition[8]

$$\mathbf{m}_0 \times \mathbf{H}_{eff}(\mathbf{m}_0) = 0 , \qquad (5)$$

which corresponds to $\partial \mathbf{m}_0 / \partial t = 0$ in the LLGE.

The magnetization deviation $\mathbf{v}$ is normal to $\mathbf{m}_0$, so that the normalization of $\mathbf{m}$ is preserved. Because of the linearity of $\mathbf{H}_{eff}$ in terms of $\mathbf{m}$, we can write $\mathbf{H}_{eff}(\mathbf{m}) = \mathbf{H}_0 + C\mathbf{v}$, where $\mathbf{H}_0 = \mathbf{H}_{eff}(\mathbf{m}_0)$. Inserting representation (4) into LLGE (1), and linearizing the equation by keeping only the terms linear in $\mathbf{v}$, $\alpha$ and $\mathbf{h}_a$, we can write a linearized LLGE for $\mathbf{v}$:

$$\frac{\partial \mathbf{v}}{\partial t} = -\gamma(\mathbf{v} \times \mathbf{H}_0 + \mathbf{m}_0 \times \mathbf{h} + \mathbf{m}_0 \times \mathbf{h}_a) + \alpha \mathbf{m}_0 \times \frac{\partial \mathbf{v}}{\partial t} , \qquad (6)$$

where $\mathbf{h} = C\mathbf{v}$ is the dynamic effective field corresponding to $\mathbf{v}$. Denoting the cross operator as $\Lambda(\mathbf{u})\mathbf{v} = \mathbf{u} \times \mathbf{v}$, we can express Eq. (6) in its matrix form

$$\frac{\partial \mathbf{v}}{\partial t} = -\gamma \left( \Lambda(\mathbf{H}_0)\mathbf{v} + \Lambda(\mathbf{m}_0)\mathbf{h} + \Lambda(\mathbf{m}_0)\mathbf{h}_a \right) + \alpha \Lambda(\mathbf{m}_0) \frac{\partial \mathbf{v}}{\partial t} . \qquad (7)$$

Since $\mathbf{v}$ is normal to $\mathbf{m}_0$, we can project every vector and operator onto the tangent space $TM(\mathbf{m}_0)$ of $\mathbf{m}_0$, by using the projection operator[9]

$$P_{\mathbf{m}_0} = (I - \mathbf{m}_0 \otimes \mathbf{m}_0) , \qquad (8)$$

where $\otimes$ is the dyadic Kronecker tensor product and $I$ is the identity matrix. It can be shown that when restricted to the vector fields in $TM(\mathbf{m}_0)$, the operator $\Lambda(\mathbf{m}_0)$ is linear and anti-symmetric and it is also invertible[9], i.e.,

$$\Lambda(\mathbf{m}_0)\Lambda(\mathbf{m}_0) = -I . \qquad (9)$$

With Eqs. (8)-(9), we can simplify Eq. (7) by multiplying by $\Lambda(\mathbf{m}_0)$ at both sides:

$$B_\perp \frac{\partial \mathbf{v}}{\partial t} + \gamma A_\perp \mathbf{v} = \gamma \mathbf{h}_a , \qquad (10)$$

where

$$\begin{aligned} A_\perp &= P_{\mathbf{m}_0}\left( (\mathbf{H}_0 \cdot \mathbf{m}_0)I - C \right) \\ B_\perp &= P_{\mathbf{m}_0}\left( \Lambda(\mathbf{m}_0) + \alpha I \right) \end{aligned} . \qquad (11)$$

Assuming the linearity of the response, the small deviation $\mathbf{v}$ can be expressed as

$$\mathbf{v} = \mathrm{Re}\{\tilde{\mathbf{v}}(\mathbf{r})e^{j\omega t}\} , \qquad (12)$$

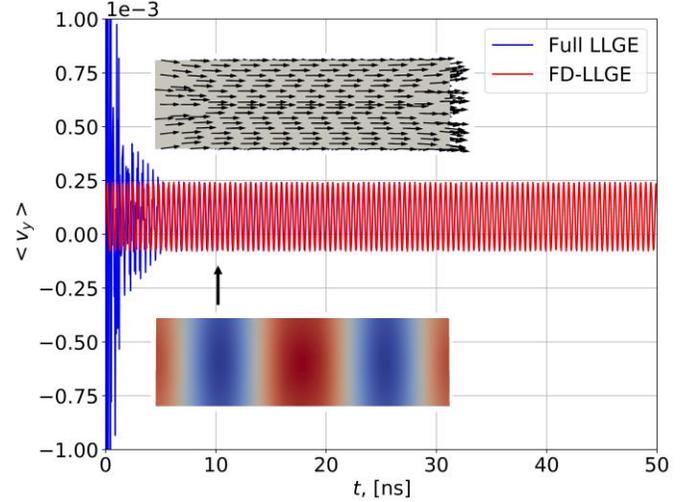

FIG. 1. Time dynamics of the average magnetization obtained via the full LLGE and FD-LLGE solvers for $L = 100\,\mathrm{nm}$, $\alpha = 0.01$. The top inset shows the strip with its magnetization equilibrium state. The bottom inset shows a snapshot of $v_y$ at $t = 10\,\mathrm{ns}$.

where $\tilde{\mathbf{v}}$ is the complex amplitude of the magnetization deviation. Inserting Eq. (12) into Eq. (10), we obtain a FD-LLGE

$$(j\omega B_\perp + \gamma A_\perp)\tilde{\mathbf{v}} = \gamma \tilde{\mathbf{h}}_a . \qquad (13)$$

Here, $\tilde{\mathbf{h}}_a$ in the right-hand side is a known function of spatial coordinates and $\tilde{\mathbf{v}}$ is an unknown, which is found by solving the linear system of equations.

The linear system matrix of Eq. (13) is dense due to the presence of the magnetostatic field operator $\mathbf{H}_{ms}$. Therefore, to enable solving large problems, iterative methods, such as the conjugated gradient (CG) or generalized minimal residual method (GMRES)[10], typically should be used. Micromagnetic problems often are stiff, which results in a badly conditioned matrix and many linear iterations required for the solution. Reducing the number of iterations can be achieved by developing a proper preconditioner. It has been shown that the high condition number for micromagnetic systems is a result of the effects of the exchange field linear operator[11]. To alleviate this problem, we use the projected sparse matrix $C_{ex\perp} = P_{\mathbf{m}_0} C_{ex}$ of the exchange field operator $C_{ex}$, which represents the discretization of the $(2A/M_s)\nabla^2$ operator in Eq. (3), to construct a sparse preconditioner matrix

$$P = (j\omega B_\perp + \gamma C_{ex\perp}) . \qquad (14)$$

The inverse matrix of preconditioner $P^{-1}$ can be approximately calculated by incomplete LU decomposition (ILU)[10] or using the block-inverse preconditioner[11]. Multiplying by $P^{-1}$ in both sides of the Eq. (13), we obtain a preconditioned linear problem

$$P^{-1}(j\omega B_\perp + \gamma A_\perp)\tilde{\mathbf{v}} = \gamma P^{-1}\tilde{\mathbf{h}}_a . \qquad (15)$$

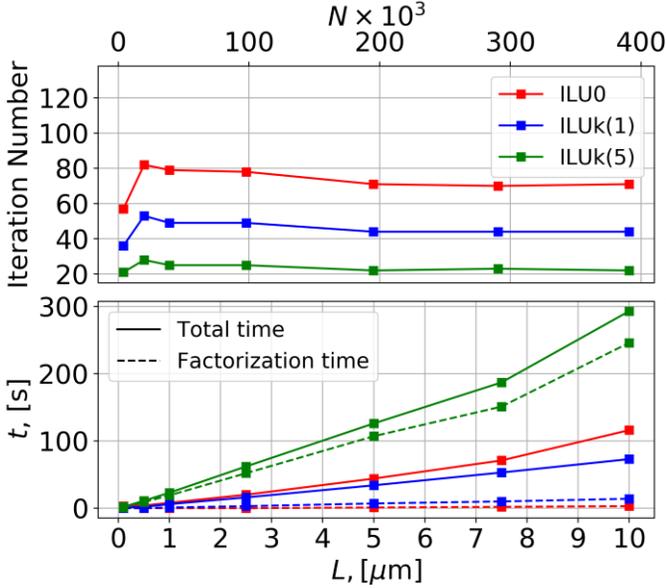

Fig. 2. Size dependence of the linear solver iteration number and computational time when using different preconditioners for $\alpha = 0.01$.

The preconditioned linear system of Eq. (17) has a significantly reduced condition number and can be solved in a much smaller number of iterations than the original problem of Eq. (13).

### III. NUMERICAL RESULTS

We implemented the FD-LLGE solver as a part of the finite element method (FEM) based micromagnetic simulator FastMag with tetrahedral elements[7]. The effective fields are computed as for the general LLGE. We note that the computation of the used FEM based effective field operator, such as Fredkin-Koehler method[12], results in non-perfectly Hermitian linear system operator $A_\perp$ [13]. Potentially, it may lead to numerical challenges if a solver assumes Hermitian matrix properties. Here, we use a solver and preconditioner that do not assume any special matrix properties. The linear system of equations (15) is solved using the GMRES algorithm with the relative error of $10^{-8}$. The preconditioner is based on the ILU0 and ILUk flavors of the ILU preconditioner using the sparse matrix $P$ of Eq. (14). The results are shown for a magnetic strip of width $w = 30\,\text{nm}$, thickness $h = 1\,\text{nm}$, and length $L > w$ ranging from 100 nm to 10 μm. The material parameters are $M_s = 800\,\text{emu/cm}^3$, $A_{ex} = 1\,\mu\text{erg/cm}$, and $\alpha$ ranges from 0.01 to 0.0001. The maximal mesh edge length was chosen as 2 nm to be sufficiently smaller than the exchange length of $\sqrt{A}/M_s = 12.5\,\text{nm}$. The equilibrium magnetization is along the longest direction (see the top inset in Fig. 1). The magnetization dynamics is excited by a weak applied magnetic field of $\tilde{\mathbf{h}}_a = 50\,\text{Oe}$ confined to the region of 10 nm near the center of the strip at frequency $f = \omega/(2\pi) = 20\,\text{GHz}$. The results were obtained on an Intel Core i9-9900K CPU.

Figure 1 demonstrates the validity of the solver by comparing the results obtained via the FD-LLGE (15) with (12) and via the original LLGE (1). The figure shows the space averaged y-component of $\mathbf{v}$ for $L = 100\,\text{nm}$. The number of mesh vertices for this case was 3,927. The number of iterations based on the ILU0 preconditioner for the FD-LLGE solver was 57 and the computational time was 3 seconds. The initial, i.e., early time dynamics, is different, which is because for the non-linear solver there are initial non-linear effects. The later time dynamics is described by the FD-LLGE solver accurately. A snapshot of $v_y$ at $t = 10\,\text{ns}$ shown in the bottom inset of Fig. 1 demonstrates a standing spin wave pattern.

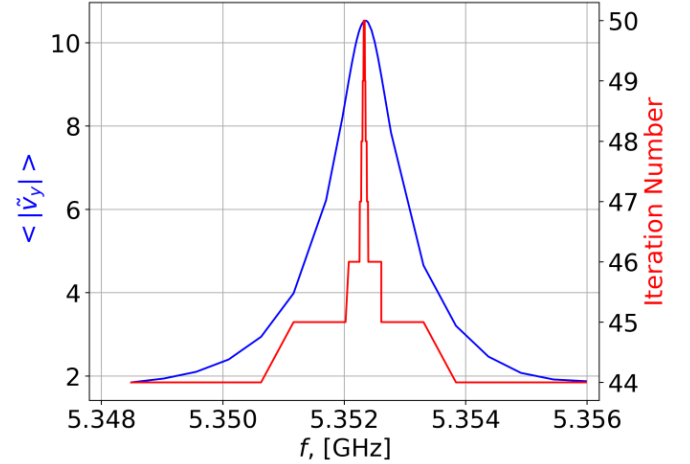

FIG. 3. Frequency dependence of the magnetization deviation magnitude for $L = 100\,\text{nm}$, $\alpha = 0.0001$.

Figure 2 demonstrates the computational performance of the FD-LLGE solver by showing the number of iterations and computational time as a function of the strip length $L$ and the corresponding number of tetrahedral mesh vertices $N$. The results are shown for the non-preconditioned formulation of Eq. (13) and preconditioned formulation of (15). It is evident that the non-preconditioned solver requires many iterations, whereas using preconditioners leads to a significant reduction of the number of iterations. The ILUk preconditioners perform better than ILU0 in terms of having a smaller number of iterations, but they have a higher cost per iterations. Overall, we find that ILUk with $k = 1$ has the best performance in terms of the computational time. Overall, the solver performance is good, and it allows addressing large scale computational problems. We note that the solving the linear system without using a preconditioner, i.e., Eq. (13), requires a large number of iterations, e.g., 1,952 iterations for the $L = 100\,\text{nm}$ case, which makes such a solver impractical. Therefore, using the preconditioned system of Eq. (15) is critical.

Finally, Fig. 3 shows the space averaged magnitude $<|\tilde{v}_y|>$ as a function of frequency $f$. One can see that the solution

$<|\tilde{v}_y|>$ exhibits a strong resonance response, with the resonant frequency related to the resonant standing spin wave excitation. The figure also shows the number of iterations with ILUk with $k=1$ preconditioner. The number of iterations increases at the resonant frequency, but this increase is modest, which demonstrates a good performance even when the excitation frequency is close to the resonant frequency.

## IV. SUMMARY AND DISCUSSION

We presented a FD-LLGE solver that allows obtaining the magnetization dynamics solutions driven by weak time harmonic excitations. The formulation is based on linearizing the original non-linear time domain LLGE and assuming a single frequency excitation and solution, which results in a single linear system of equations for the complex magnetization deviation around the magnetization equilibrium state. The linear system is preconditioned by a sparse preconditioner that allows significantly reducing the number of iterations and computational time required for its solution.

We note that once the solutions for $\tilde{\mathbf{v}}$ is obtained, the time domain solution $\mathbf{v}$ is found for all time via Eq. (12).

Assuming a small number of linear iterations for solving Eq. (15), the FD-LLGE solver provides a much more efficient approach than solving the original non-linear LLGE for finding solutions in the linear regime. If the excitation is given by multiple frequencies, e.g., by a pulse, then multiple frequency domain solutions can be combined via the Fourier transform. The results are shown for real-valued frequencies, but solutions can also be obtained for complex-valued frequencies, which can provide insights into the magnetization dynamics behavior. The FD-LLGE solver is implemented in the FEM framework, and it can also be extended to finite difference implementations.

## ACKNOWLEDGMENTS

This work was supported as part of Quantum Materials for Energy Efficient Neuromorphic-Computing (Q-MEEN-C), an Energy Frontier Research Center funded by the U.S. Department of Energy, Office of Science, Basic Energy Sciences under Award No. DESC0019273. This work used the XSEDE[14], which is supported by NSF Grant # ACI-1548562, specifically, it used the Bridges and Comet systems supported by NSF Grant # ACI-1445506.

## DATA AVAILABILITY STATEMENT

The data that support the findings of this study are available from the corresponding author upon reasonable request.